\documentclass[english]{article}
\usepackage[T1]{fontenc}
\usepackage[latin9]{inputenc}
\usepackage{geometry}
\usepackage{babel}
\usepackage{array}
\usepackage{amscd}
\usepackage{amssymb}
\usepackage{amsmath}
\usepackage{mathtools}
\usepackage{multirow}
\usepackage{enumitem}
\usepackage{graphicx}
\usepackage{color}
\usepackage{authblk}
\usepackage{subcaption}
\usepackage{csquotes}
\usepackage[colorlinks]{hyperref}
\usepackage[sorting=none]{biblatex}

\begin{document}

\title{Generalized Hall current on a finite lattice}

\author{
  Srimoyee Sen, Semeon Valgushev \\
  \text{Department of Physics and Astronomy, Iowa State University, Ames, IA, 50011} 
}

\maketitle
\begin{abstract}
Gapped fermion theories with gapless boundary fermions can exist in any number of dimensions. When the boundary has even space-time dimensions and hosts chiral fermions, a quantum Hall current flows from the bulk to the boundary in a background electric field. This current compensate for the boundary chiral anomaly. Such a current inflow picture is absent when the boundary theory is odd dimensional.
However, in recent work, the idea of quantum Hall current has been generalized to describe odd dimensional boundary theories in continuous Euclidean space-time dimension of infinite volume. In this paper we extend this idea to a lattice regulated finite volume theory of 1+1 dimensional Wilson-Dirac fermions. This fermion theory with a domain wall in fermion mass can host gapless modes on the wall. The number of gapless fermions is equal to the integral of the divergence of the lattice generalized Hall current.

\end{abstract}
\section{Introduction}
Odd dimensional Dirac fermion field theories are interesting when there is a domain wall in fermion mass. In that case, the domain wall defect is even dimensional and hosts massless chiral fermions \cite{Callan:1984sa}. When this theory is coupled to electromagnetic fields, the boundary suffers from chiral anomaly leading to non-conservation of vector current in the presence of background electromagnetic fields. However, as Callan-Harvey showed \cite{Callan:1984sa}, a vector current flows from the bulk to the boundary restoring current conservation in the higher dimensional theory. In order to compute this current one integrates out the fermion away from the domain wall which leaves behind a Chern-Simons theory for the electromagnetic field. This explains the inflowing current from the bulk to the boundary. As is well known, the odd dimensional gapped bulk theory of free Dirac fermion describes the physics of quantum Hall effect. The inflowing current is analogous to the quantum Hall current whereas the massless chiral fermions on the domain wall are analogs of the quantum Hall edge states. 

More generally, gapped fermion field theories can host massless fermions on domain walls irrespective of whether the wall is even or odd dimensional. They describe the physics of topological insulators and superconductors with corresponding edge states in various dimensions \cite{Kitaev:2009mg, Ryu_2010, PhysRevB.85.085103, PhysRevB.78.195125,  Ludwig_2015}. When the boundary is odd dimensional, in contrast to quantum Hall effect, the boundary theory does not have a chiral anomaly. Therefore, we don't expect an inflowing current from the bulk to the boundary as in the case of quantum Hall effect. Although, the boundary theory can have discrete anomalies which connects the existence of the edge states to the gapped bulk theory\cite{Witten:2015aba}. In a recent paper \cite{kaplan2021index, Kaplan:2022uoo} the authors showed that the idea of the Hall current can be generalized to odd dimensional boundaries. The idea was inspired by index calculation of fermion vortex system in \cite{PhysRevD.24.2669}.
This generalization of the Hall current relies on the following step:
the Minkowski space domain wall fermion theory with massless boundary fermion is first connected to another Euclidean fermion theory where the Euclidean fermion operator has a nonzero index. This index equals the number of massless fermions in the original Minkowski theory.
From there, it was shown \cite{kaplan2021index, Kaplan:2022uoo} that one can construct a generalized Hall current: the space-time integral of its divergence equaling the index of the fermion operator. The construction outlined in \cite{kaplan2021index, Kaplan:2022uoo} holds for non-interacting fermions in infinite volume and continuum space-time. The goal of this paper is to extend that analysis to a discrete space-time lattice of infinite and finite volume. The analysis in \cite{kaplan2021index, Kaplan:2022uoo} included several different fermion theories in various space-time dimensions. In this paper, we choose to work with the simplest example: $1+1$ dimensional Dirac fermion with a domain wall in its mass \cite{PhysRevD.13.3398}. The domain wall hosts a massless fermion which may suffer from discrete anomalies (\cite{ Witten:2015aba, PhysRevB.83.075103}), but does not suffer from chiral anomaly. As a result, one doesn't expect a Hall current flowing from bulk to the boundary. However, the generalized Hall current exists for this system in infinite volume and continuum space-time. We explore how the generalized Hall current for this system can be constructed on an infinite and finite lattice.
 
A crucial observation which makes the continuum construction of the generalized Hall current possible is the following. Whenever the index of a Euclidean elliptic fermion operator is nonzero, there is a current in the system: the space-time integral of the divergence of this current equals the index. We call this current the generalized Hall current. Note that the index of a Euclidean elliptic operator is the difference between the number of zero modes of that operator and that of its Hermitian conjugate \cite{PhysRevD.24.2669}. Therefore, no generalized Hall current exists if the index of the operator is zero.  This observation is not meant to be self-evident and its proof is outlined in \cite{kaplan2021index, Kaplan:2022uoo}. We will discuss the proof briefly in the next section of this paper. This observation can then be used to construct the generalized Hall current for massless fermion edge states of any Minkowski fermion theory as follows. The first step is to use the Minkowski fermion operator to construct its Euclidean counterpart. Since the Minkowski fermion operator has massless states living on the defect, the corresponding Euclidean operator has unnormalizable zero eigenvalue eigenstates living on the same defect. These states are not zero modes since they are not normalizable. As a result the index of the Euclidean operator at this stage is zero. Ref. \cite{kaplan2021index, Kaplan:2022uoo} then introduces a slight deformation to this Euclidean operator through the introduction of a background diagnostic field in such a way that this unnormalizable zero eigenvalue state becomes localized and normalizable. 
i.e. the deformed Euclidean operator has a zero mode iff the original Minkowski fermion operator had a massless fermion in its spectrum. The introduction of this diagnostic field also creates an imbalance between the number of zero modes of the fermion operator and its Hermitian conjugate resulting in a nonzero index for the deformed theory. Additionally, the construction carefully ensures that the index survives in the limit of the diagnostic field being taken to zero. We expect a generalized Hall current to flow as long as the index is nonzero. In the continuum analysis, one can obtain this Hall current by simply perturbing in the diagnostic field and integrating out the fermions in a one loop diagram. This is analogous to the Goldstone-Wilczek calculation \cite{PhysRevLett.47.986}.

As we embark on generalizing the above construction on the lattice, both infinite and finite, we explore which elements of the continuum construction can be carried over to the lattice without significant modification and which elements need to be reformulated. Since we will work with the $1+1$ dimensional fermion theory, from this point onward we exclusively focus on it. The organization of the paper is as follows. We will begin with a brief overview of the generalized Hall current construction in the continuum specializing to the case in $1+1$ dimensions. We will then discuss how this construction is generalized to an infinite lattice analytically. The following section will describe the numerical analysis of this construction and demonstrate that a generalized Hall current exists on a finite lattice.

\section{Infinite volume continuum analysis}
The procedure for constructing the generalized Hall current in the continuum in infinite volume is described in detail in \cite{kaplan2021index, Kaplan:2022uoo}. We briefly review this construction here. Consider a Minkowski fermion operator $D_M$ with a mass defect which causes it to have a massless fermion in the spectrum that is stuck to the defect.  To construct the generalized Hall current 
\begin{enumerate}
    \item We analytically continue this fermion operator to Euclidean space-time, denoting it by $\mathcal{D}$.
    \item Introduce background diagnostic field to deform the fermion operator $\mathcal{D}$ to have an index of one (equal to the number of massless fermions in the original Minkowski theory). 
    \item Obtain the generalized Hall current following a Goldstone-Wilczek \cite{PhysRevLett.47.986} inspired calculation using one loop Feynman diagram. 
    \item Take the background diagnostic field to zero at the end of calculation and confirm that the generalized Hall current and the index survives taking this limit. 
\end{enumerate}

Before we apply this construction to $1+1$ dimensional example, let's first attempt to understand how the index of a fermion operator gives rise to an inflowing current in infinite volume  continuous space-time of Euclidean signature. Note that the index of the fermion operator $\mathcal{D}$ is given by 
\begin{eqnarray}
I=\text{Dim}(\text{ker}\,D)-\text{Dim}(\text{ker}\,D^{\dagger}).
\end{eqnarray}
In the example we will consider, the number of zero modes of either the operator $\mathcal{D}$ or the operator $\mathcal{D}^{\dagger}$ is zero. As a result the magnitude of the index ends up being equal to the number of zero modes of one operator or the other. Furthermore, the number of zero modes of the operator $\mathcal{D}$ coincides with the number of zero modes for the operator $\mathcal{D}^{\dagger}\mathcal{D}$ and the number of zero modes of $\mathcal{D}^{\dagger}$ coincides with that of $\mathcal{D}\mathcal{D}^{\dagger}$. Therefore the formula for the index can be re-expressed by defining 
\begin{eqnarray}
\mathcal{I}(M)=\frac{M^2}{M^2+\mathcal{D}^{\dagger}\mathcal{D}}-\frac{M^2}{\mathcal{D}\mathcal{D}^{\dagger}+M^2}
\end{eqnarray}
and noting that 
\begin{eqnarray}
I=\lim_{M\rightarrow 0} \mathcal{I}(M). 
\end{eqnarray}
Interestingly, the quantity $\mathcal{I}(M)$ can now be recast as the matrix element $\mathcal{I}(M)=-\int d^{d+1}x\langle\bar{\Psi}\Gamma_\chi\Psi\rangle$ in a fermion theory with the following action 

\begin{eqnarray}
\mathcal{S}=\int d^{d+1}x\, \bar{\Psi}(K+M)\Psi
\label{theory}
\end{eqnarray}
where 
\begin{eqnarray}
\label{eq:doubled_K}
K=\begin{pmatrix}
0 && -\mathcal{D}^{\dagger}\\
\mathcal{D} && 0
\end{pmatrix}
\end{eqnarray}
and 
\begin{eqnarray}
\label{eq:doubled_gamma_5}
\Gamma_\chi=\begin{pmatrix}
1 && 0\\
0 && -1
\end{pmatrix}.
\end{eqnarray}
Note that, $d+1$ is the number of space-time dimensions in which the original fermion operator $\mathcal{D}$ is defined. The spinor $\Psi$ has twice the dimension of the spinors of the original theory. The gamma matrices for this theory can be easily read off using 
\begin{eqnarray}
\Gamma_{\mu}=i\partial \tilde{K}(p)/\partial p_{\mu}
\end{eqnarray}
where $\tilde{K}$ is the Fourier transform of $K$.
The theory of Eq. \ref{theory} has its own fermion number symmetry which works as $\Psi\rightarrow e^{i\theta}\Psi$. In the $M\rightarrow 0$ limit, it also has an axial symmetry $\Psi\rightarrow e^{i\Gamma_{\chi}\alpha}\Psi$ where this new axial symmetry has nothing to do with the symmetries of the original theory. We can now construct an axial current $\mathcal{J}_{\mu}^{\chi}=\bar{\Psi}\Gamma_{\mu}\Gamma_{\chi}\Psi$ and write down the Ward identity for it 
\begin{eqnarray}
\partial_{\mu}\mathcal{J}_{\mu}^{\chi}=2M\bar{\Psi}\Gamma_{\chi}\Psi-\mathcal{A}
\label{eqJ}
\end{eqnarray}
where $\mathcal{A}$ is the ``anomaly contribution" 
\begin{eqnarray}
\mathcal{A}=-2 \lim_{\Lambda\rightarrow\infty} \text{Tr}(\Gamma_{\chi}e^{K^2/\Lambda^2})=-2\mathcal{I}(\infty).
\end{eqnarray}
This anomaly contribution can be computed using the methods outlined in Fujikawa \cite{PhysRevLett.42.1195}. It is found to vanish for the theory under consideration Eq.~\ref{theory} and was elaborated in \cite{kaplan2021index, Kaplan:2022uoo}. At this point we can take the limit $M\rightarrow 0$ in Eq.~\ref{eqJ} to write 
\begin{eqnarray}
\label{eq:index_vs_divj}
I=\mathcal{I}(0)=-\lim_{M\rightarrow 0} M\int d^{d+1}x \langle\bar{\Psi}\Gamma_{\chi}\Psi\rangle=-\lim_{M\rightarrow 0}\frac{1}{2}\int d^{d+1}x \langle \partial_{\mu}\mathcal{J}_{\mu}^{\chi} \rangle.
\end{eqnarray}
We have now expressed the index of the fermion operator in terms of the ``axial" current of the theory in Eq.~\ref{theory}. We call this current the generalized Hall current. This generalized Hall current $\bar{\Psi}\Gamma_{\mu}\Gamma_{\chi}\Psi$ can now be computed using one loop Feynman diagrams by perturbing in the mass defect as well as the other background fields. We will review how this is done for $1+1$ dimensional Dirac fermion with a domain wall in its mass.

\subsection{$1+1$ dimensional Dirac fermion in continuum}
Let's consider the Lagrangian of a Dirac fermion in Minkowski space-time with Dirac mass denoted as $\phi_1$. It has the Lagrangian
\begin{eqnarray}
\mathcal{L}=\bar{\psi}(i\gamma^{\mu}\partial_{\mu}-\phi_1)\psi
\label{eq:minkowski_lag}
\end{eqnarray}
where $\mu$ takes values $0$ and $1$, $x_0$ is the temporal and $x_1$ is the spatial coordinate.
We can take the $\gamma$ matrices as 
\begin{eqnarray}
\gamma^0=\sigma_2, \gamma^1=-i\sigma_1, \gamma^{\chi}=\sigma_3
\end{eqnarray}
where $\gamma_{\chi}$ is the chirality operator. 
If we introduce a domain wall in $\phi_1$ along the spatial coordinate $x_1$, $\phi_1=m_0\epsilon(x_1)$ with $m_0>0$ and
\begin{eqnarray}
\label{eq:epsilon_function}
\epsilon(x) = 
\begin{cases}
    +1, & x \geq 0\\
    -1, & x < 0
\end{cases},
\end{eqnarray}
then we will have a massless fermion mode living on the domain wall at $x_1=0$
as seen from the Dirac equation in the domain wall background
\begin{eqnarray}
i\gamma^0\partial_0\psi+i\gamma^1\partial_1\psi-\phi_1\psi=0. 
\end{eqnarray}
To look for massless state, we can set $\partial_0\psi=0$ and find that the Dirac equation is solved by 
\begin{eqnarray}
\psi=\frac{1}{\sqrt{2}}\begin{pmatrix}1\\
-1
\end{pmatrix}e^{-m_0 |x_1|}.
\end{eqnarray}
In order to construct the generalized Hall current we first have to analytically continue to Euclidean space-time where the Lagrangian is now
\begin{eqnarray}
\mathcal{L}_\text{E}=\bar{\psi}(\gamma_{\mu}\partial_{\mu}+\phi_1)\psi
\end{eqnarray}
with Euclidean gamma matrices defined as
\begin{eqnarray}
\gamma_0=\sigma_2, \gamma_1=-\sigma_1, \gamma_\chi=\sigma_3.
\end{eqnarray}
We also denote two dimensional identity matrix as $\sigma_0$.
The corresponding fermion operator $\gamma_{\mu}\partial_{\mu}+\phi_1$ has an unnormalizable zero eigenvalue eigenstate. However this state doesn't count as zero mode which should be normalizable. In order to engineer a zero mode we turn on a background pseudo-scalar field with a domain wall profile in the Euclidean time direction. We also refer to this field as a diagnostic field. The corresponding Lagrangian is of the form 
\begin{eqnarray}
\label{eq:euclid_lag}
\mathcal{L}_\text{E}=\bar{\psi}(\gamma_{\mu}\partial_{\mu}+\phi_1+i\phi_2\gamma_\chi)\psi
\end{eqnarray}
where $\phi_2=\mu_0\epsilon(x_0)$ with $\mu_0>0$.
Let's denote this fermion operator as $\mathcal{D}$ with
\begin{eqnarray}
\mathcal{D}=(\gamma_{\mu}\partial_{\mu}+\phi_1+i\phi_2\gamma_\chi).
\label{Df}
\end{eqnarray}

We find that the operator $\mathcal{D}$ has one zero mode of the form
\begin{eqnarray}
\psi=\frac{1}{\sqrt{2}}\begin{pmatrix}1 \\ -1\end{pmatrix}e^{-m_0|x_1|-\mu_0|x_0|}. 
\end{eqnarray}
We can also look for zero modes for the operator $\mathcal{D}^{\dagger}$ and find that there are none for this specific choice of domain wall profile ($m_0>0, \mu_0>0$). More generally, for other choices of the domain wall profile, e.g. with $m_0>0, \mu_0<0$ or $m_0<0, \mu_0>0$ we find a zero mode for the operator $\mathcal{D}^{\dagger}$ and the operator $\mathcal{D}$ has no zero modes. Similarly, the choice of $m_0<0, \mu_0<0$ yields a zeromode for $\mathcal{D}$ and none for $\mathcal{D}^{\dagger}$. In other words, the magnitude of the index of the fermion operator remains $1$ as long as there is a domain wall in both $\phi_1$ and $\phi_2$. However, whether the index is positive or negative depends on the profile of choice. 

There is a simple way to relate the domain wall profile with the index of the fermion operator. To see this, we can first express $\phi_1+i\phi_2$ as $\phi_1+i\phi_2=v e^{i\theta}$. It is easy to see that for a crossed domain wall profile in $\phi_1$ and $\phi_2$, if one considers a polar coordinate system centered at $x_0=x_1=0$, then the phase variable $\theta$ completes a winding of $2\pi$ or $-2\pi$ as one travels along a contour encircling the center over a polar angle of $2\pi$. The crossed domain wall defect can therefore be thought of as a vortex in $\phi_1+i\phi_2$.
We have now constructed the intended fermion operator whose index is equal to the winding in the crossed domain wall configuration. Note that the index and the winding survives in the limit $\mu_0\rightarrow 0$.

\subsection{Generalized Hall current (GHC) in the continuum}
We now review the one loop Feynman diagram calculation to compute the generalized Hall current and then verify that the space-time integral of its divergence equals the index. 

Following the prescription outlined in Eq. \ref{theory},\ref{eq:doubled_K},\ref{eq:doubled_gamma_5} we construct the $K$ matrix which we can re-express in momentum space as 
\begin{eqnarray}
K=\Gamma_{\mu}k_{\mu}+i \phi_2\Gamma_2+i\phi_1\Gamma_3
\end{eqnarray}
where we have defined 
\begin{eqnarray}
\Gamma_i=\sigma_1\otimes\gamma_i, \Gamma_2=\sigma_1\otimes\gamma_\chi, \nonumber\\
\Gamma_3=-\sigma_2\otimes\sigma_0, \Gamma_{\chi}=\sigma_3\otimes\sigma_0.
\end{eqnarray}
To compute the ``axial" current, we rewrite the mass terms as $\phi_1+i\phi_2=(v+\rho(x))e^{i\theta(x)}$ and expand the $K$ matrix in $\theta$
with
\begin{eqnarray}
K=K_0+\delta K
\end{eqnarray}
where $K_0=\gamma_{\mu}k_{\mu}+\rho$ and $\delta K=i v\theta \Gamma_2+i\rho\Gamma_3$. Up to linear order in $\theta$ we get
\begin{eqnarray}
\mathcal{J}_{\mu}^{\chi}&=&+v\frac{\partial\theta}{\partial x_{\mu}}\int \frac{d^2q}{(2\pi)^2}\text{Tr}\left(\Gamma_{\mu}\Gamma_{\chi}\frac{dK_0^{-1}}{dq_{\nu}}\Gamma_2 K_0^{-1}\right)\nonumber\\&=&\epsilon_{\mu\nu}\partial_{\nu}\theta\int \frac{d^2q}{(2\pi)^2}\frac{4v^2}{(q^2+v^2)^2}\nonumber\\
&=& \frac{1}{\pi}\epsilon_{\mu\nu}\partial_{\nu}\theta
\end{eqnarray}
We can now compute the space-time integral of the divergence
of this current and relate it to the index with  
\begin{eqnarray}
\mathcal{I}(0)=-\frac{1}{2}\int d^2x \partial_{\mu}\mathcal{J}_{\mu}^\chi=-\nu_{\theta}
\end{eqnarray}
where $\nu_{\theta}$ is the winding of the crossed domain wall or vortex configuration. For the specific domain wall profile we have chosen this winding is $-1$. Therefore we get an index of $1$ which is consistent with the index we obtained for the fermion operator in the previous subsection. 
This demonstrates that whenever the Minkowski theory specified by Eq.~\ref{eq:minkowski_lag} has a domain wall in fermion mass hosting massless edge state, one can construct a corresponding Euclidean fermion operator with the following properties:
\begin{enumerate}
    \item The Euclidean fermion operator has an index of $\pm 1$ in the presence of a background diagnostic field.
    \item In the limit of diagnostic field going to zero this Euclidean operator coincides with the Euclidean analytic continuation of the Minkowski operator in Eq.~\ref{eq:minkowski_lag}.
    \item The index of this Euclidean operator persists in the limit of the diagnostic field being taken to zero and is equal to the space-time integral of the divergence of the GHC.
\end{enumerate}
In the next section, we will to extend our Euclidean fermion operator construction to discrete space-time. In order to mimic the continuum construction sufficiently closely we will have to maintain the following
\begin{enumerate}
    \item The lattice fermion operator or its Hermitian conjugate should not have more than one zeromode. 
    \item We will exclude regions in parameter space where the number of zeromodes for the fermion operator and its Hermitian conjugate are the same. 
\end{enumerate}
The second condition ensures that the index of the fermion operator is nonzero.

\section{1+1 case on the lattice in infinite volume}
\label{lat}

We begin with the fermion operator in Eq. \ref{Df} and discretize spacetime setting the lattice spacing to $1$. If we first set $\phi_2=0$ and naively discretize space-time, we observe an important difference from the spectrum in the continuum : i.e. we see fermion doubling. 
This is to say, in the continuum we had a single solution to the equation $\mathcal{D}|_{\phi_2=0}\psi=0$ with $\psi$ being localized in the $x_1$ direction and constant in the $x_0$ direction. 
On the lattice, there are more than one solution of this form. In order to remove fermion doubling so as to retain only one solution will require us to introduce higher dimensional operators to the Lagrangian similar to the Wilson terms used in domain wall fermions \cite{Kaplan:1992bt, Jansen:1992yj, Golterman:1992ub, Jansen:1992tw, Sen:2022dkl}. Since our end goal is to construct a Euclidean fermion operator with a single zeromode we have two simple choices for this higher dimensional term:

\begin{enumerate}
    \item {\bf {Wilson-like operator:}} Inspired by the Wilson term in lattice field theory, we introduce the following higher-derivative operators to the Lagrangian which we call Wilson-like terms \cite{Sen:2022dkl}:
   \begin{eqnarray}
    \label{eq:op_wilson_like}
    \mathcal{D}_1 = \sum_\mu \gamma_\mu \nabla_\mu + \frac{R}{2}\nabla_1^2+i \gamma_\chi \frac{R}{2}\nabla_0^2.
    \end{eqnarray}
    We set parameter $R=1$.
    \item {\bf{Fermion operator with Wilson term:}} We introduce in the Lagrangian the standard Wilson term:
   \begin{eqnarray}
    \label{eq:op_wilson_dirac}
    \mathcal{D}_2 = \sum_\mu \gamma_\mu \nabla_\mu + \frac{R}{2}\sum_{\mu}\nabla_{\mu}^2
    \end{eqnarray}
    We again set Wilson parameter to $R=1$.
\end{enumerate}
We now look for the zeromodes of these operators by varying the paramaters of our theory. 
\subsection{Zeromodes}
In this subsection we aim to obtain zeromode solutions, by varying the parameters like the domain wall heights for the two types of lattice fermion operator introduced in the previous section. We first present an analytic calculation for the zeromode of the Wilson-like operator in infinite and finite volume. The corresponding expressions for the zeromode profile are simple and illuminating. An analogous analytic calculation for the Wilson fermion case is more difficult and not particularly illuminating. Therefore we defer the discussion of the Wilson fermion operator to the subsection \ref{subnum} where we present numerical analysis of both the Wilson fermion case and the Wilson-like cases.

\subsubsection{Analytic solution for the zeromode in infinite volume}
\label{sec:wilson_like_op}

We begin with Wilson-like operator given by Eq.~\ref{eq:op_wilson_like}:
\begin{eqnarray}
\mathcal{D}_1=\begin{pmatrix}
\tilde{\phi}_1+i\tilde{\phi}_2 && -i\nabla_0-\nabla_1\\
i\nabla_0-\nabla_1 && \tilde{\phi}_1-i\tilde{\phi}_2
\end{pmatrix}
\end{eqnarray}
where $\tilde{\phi}_1=\phi_1+\frac{1}{2}\nabla_1^2$ and $\tilde{\phi}_2=\phi_2+\frac{1}{2}\nabla_0^2$. 
With an ansatz of $\psi_+=\begin{pmatrix} 1 \\ -1\end{pmatrix}\varphi_+$ of $\gamma_1$ eigenvalue $+1$ we get two equations for $\varphi_+$,
\begin{eqnarray}
\nabla_1\varphi+\tilde{\phi}_1\varphi_+=0,\\
\label{e21}
\nabla_0\varphi+\tilde{\phi}_2\varphi_+=0.
\label{e22}
\end{eqnarray}
Then using an ansatz of $\varphi_+=z_0^{x_0}z_1^{x_1}$ we see that there exists normalizable solution with 
\begin{eqnarray}
\label{eq:zm_wilson_like}
z_0=(1-\phi_2)\nonumber\\
z_1=(1-\phi_1)
\end{eqnarray}
when $0<m_0<2$ and $0<\mu_0<2$. Let's fix $m_0=1$ and $\mu_0=1$. 
Now we consider the ansatz of $\psi_-=\begin{pmatrix} 1 \\ 1\end{pmatrix}\varphi_-$. The EOMs for $\varphi_-$ are
\begin{eqnarray}
\nabla_1\varphi_--\tilde{\phi}_1\varphi_-=0,\\
\label{e11}
\nabla_0\varphi_--\tilde{\phi}_2\varphi_-=0.
\label{e12}
\end{eqnarray}
These are solved by the ansatz 
\begin{eqnarray}
\varphi_-=z_0^{x_0} z_1^{x_1} \,\,\,\, \text{with}\,\,\,\,
z_0=\frac{1}{(1-\phi_2)}, \,\,\, z_1=\frac{1}{(1-\phi_1)}.
\end{eqnarray}
The solution is not normalizable for our choice of $m_0=1$ and $\mu_0=1$. Therefore $\psi_-$ is not a zeromode of $\mathcal{D}_1$; thus $\mathcal{D}_1$ has a single zeromode specified by the expression for $\psi_+$ in Eq.~\ref{eq:zm_wilson_like}. 
Now, let's look at the zero modes for $\mathcal{D}^{\dagger}$. With an ansatz of $\xi_-=\begin{pmatrix} 1 \\ 1\end{pmatrix}\chi_-$ and $\xi_+=\begin{pmatrix} 1 \\ -1\end{pmatrix}\chi_+$ we get the following EOMs for $\chi_-$ and $\chi_+$,
\begin{eqnarray}
\nabla_1\chi_-+\tilde{\phi}_1\chi_-=0\nonumber\\
\nabla_0\chi_--\tilde{\phi}_2\chi_-=0
\label{e31}
\end{eqnarray}
and
\begin{eqnarray}
\nabla_1\chi_+-\tilde{\phi}_1\chi_+=0\nonumber\\
\nabla_0\chi_++\tilde{\phi}_2\chi_+=0
\label{e32}
\end{eqnarray}
Using an ansatz of the form $z_0^{x_0}z_1^{x_1}$ for $\chi_-$ and $\chi_+$ we see that there are no normalizable solutions for either. 
Thus we have accomplished what we set out do, i.e. engineer a Euclidean fermion operator on the lattice with an index of $+1$ using the Wilson-like terms.

Note that, if we vary parameters the pattern of zeromodes change. E.g. for $-2<m_0<0, -2<\mu_0<0$ we find a zeromode solution with $\gamma_1$ eigenvalue $-1$. Similarly, with $2>m_0>0, 0>\mu_0>-2$ and $0>m_0>-2, 2>\mu_0>0$ we find no normalizable zeromode for the operator $\mathcal{D}_1$. However, we find a zeromode for the operator $\mathcal{D}_1^{\dagger}$: $\gamma_1$ eigenvalue $-1$ for  $2>m_0>0, 0>\mu_0>-2$ and $\gamma_1$ eigenvalue $1$ for $2>\mu_0>0, 0>m_0>-2$.

\subsubsection{Finite volume}
Our next goal is to generalize the infinite volume construction to finite volume, i.e. on $S^1 \times S^1$. At this point we will have to resort to numerical techniques. We will take the lattice size to be $L\times L$ where the domain wall in $\phi_1$ is located at $x_1=0$ and the anti-wall is located at $x_1=L/2$. Similarly, the domain wall in $\phi_2$ is located at $x_0=0$ with anti-wall at $x_0=L/2$. Therefore in effect we have four vortex-like defects 
at $(x_0=0, x_1=0)$, $(x_0=0, x_1=L/2)$, $(x_0=L/2, x_1=0)$ and $(x_0=L/2, x_1=L/2)$. 
There are several subtleties with this finite volume analysis which we describe below.

\paragraph{Exact zeromode and tuning:} The two types of lattice fermion operators, which we call the Wilson-like or Wilson fermion operator, will in general not exhibit exact zeromodes in finite volume for arbitrary choice of domain wall heights. 
To understand why this is the case, consider the Wilson-like fermion operator. 
Since we are considering $S^1\times S^1$ with periodic boundary condition, any solution to the equation of motion including the zeromode should satisfy:
\begin{eqnarray}
\label{eq:_wilson_like_pbc}
\phi_+(x_\mu = -L/2) = \phi_+(x_\mu = L/2)
\end{eqnarray}
for $\mu=0,1$. The solution obtained in Eq.~\ref{eq:zm_wilson_like} for an infinite lattice with equal magnitude of domain wall height on the two sides of the wall will not satisfy this periodic boundary condition (PBC) in finite volume. 
In order to obtain an exact zeromode solution which satisfies PBC, we will need to assume more general domain wall configuration:
\begin{eqnarray}
\label{eq:domain_wall_generic}
\phi_1(x_1)&=&\begin{cases}
      m_+ & x_1\geq 0\\
      m_- & x_1<0\\
    \end{cases},\\ \nonumber
\phi_2(x_0)&=&\begin{cases}
      \mu_+ & x_0\geq 0\\
      \mu_- & x_0<0\\
    \end{cases}.
\end{eqnarray}
Then we find an exact zeromode for the choice
\begin{eqnarray}
\label{eq:tuning_wilson_like_pbc}
\frac{1}{1-m_-} = (1-m_+),\\ \nonumber
\frac{1}{1-\mu_-} = (1-\mu_+).
\end{eqnarray}
Note that these equations do not depend on the lattice size, thus if they are satisfied then the exact zeromode of $\mathcal{D}_1$ will exist in any volume. A similar analysis is much more complicated for the Wilson case and is not particularly interesting. 

It's important consider however, that the Minkowski space domain wall theory in continuous space-time and in infinite volume hosts massless edge states without requiring any tuning of the domain wall height. Therefore, on the finite lattice too, we seek a formulation which does not rely on tuning of the domain wall heights. Since, a finite volume lattice fermion operator $\mathcal{G}$ does not have an exact zeromode in general, we shift our attention to the operator $\mathcal{G}^{\dagger}\mathcal{G}$. This is also motivated by the observation that the index formula for the fermion operator involves the kernel of the operator $\mathcal{G}^{\dagger}\mathcal{G}$ and $\mathcal{G}\mathcal{G}^{\dagger}$.
However, the operator $\mathcal{G}^{\dagger}\mathcal{G}$ (or $\mathcal{G}\mathcal{G}^{\dagger}$) doesn't have exact zeromodes in finite volume either. In order to recover them one has to take an infinite volume limit. Interestingly, this limit is smooth for $\mathcal{G}^{\dagger}\mathcal{G}$ (or $\mathcal{G}\mathcal{G}^{\dagger}$) but not necessarily for $\mathcal{G}$ (or $\mathcal{G}^{\dagger}$) itself.
We will use this observation to enable the GHC construction.
The index formula in infinite volume is related to the
the difference of the zeromodes of the operators $\mathcal{D}_{1/2}\mathcal{D}_{1/2}^{\dagger}$ and $\mathcal{D}_{1/2}^{\dagger}\mathcal{D}_{1/2}$. We will work with the same definition for the ``index'' in finite volume. 
As we will see, in finite volume, the operators, $\mathcal{D}_{1/2}\mathcal{D}_{1/2}^{\dagger}$ and $\mathcal{D}_{1/2}^{\dagger}\mathcal{D}_{1/2}$ will exhibit smooth convergence towards infinite volume zeromodes without any fine tuning for the domain wall heights, whereas $\mathcal{D}_{1/2}$ will not. This will enable us to construct a tuning independent lattice GHC.
Although we don't need fine tuning of domain wall heights, the domain wall heights must satisfy the following constraints to host a zeromode in the infinite volume limit. E.g. for a crossed domain wall configuration of the form $\phi_1=m_0\epsilon(x_1)$ and $\phi_2=\mu_0\epsilon(x_0)$ we must have $0<m_0,\mu_0<2$ in order for there to be a zeromode. Therefore, in the rest of the paper we will choose parameters that satisfy this condition.
Finally, even though our goal is to construct a GHC formulation which does not rely on tuning of the domain wall height, we will present the results for the tuned case of the Wilson-like fermion operator to illustrate a GHC in the presence of an exact lattice zeromode.

\paragraph{Index in finite volume:}
In a finite volume, a domain wall setup will appear accompanied by an anti-wall. As a result, with a domain wall in mass and the diagnostic field, we will have four vortex, two vortex and two anti-vortex defects in finite volume as described in the beginning of this subsection. Clearly the net winding of this system is zero. Therefore the net ``index" in this finite volume lattice theory is also zero.  However, locally in a region near each of the vortex defect we should be able to define an "index" which we can then attempt to connect to a lattice version of the generalized Hall current. 
In other words, in finite volume, the  operators $\mathcal{D}_{1/2}\mathcal{D}_{1/2}^{\dagger}$ and $\mathcal{D}_{1/2}^{\dagger}\mathcal{D}_{1/2}$ have the same number of zeromodes. This implies that the difference between the number of zeromodes for the two is zero, or the net ``index" is zero. 
However, the zeromodes for these two operators will be localized on different vortex defects. E.g. $\mathcal{D}_{1}^{\dagger}\mathcal{D}_{1}$ will have a zeromode on the defect at $(x_0=0, x_1=0)$ and $(x_0=L/2, x_1=L/2)$. Similarly,  $\mathcal{D}_{1}\mathcal{D}_{1}^{\dagger}$ will have zeromodes at
$(x_0=L/2, x_1=0)$ and $(x_0=0, x_1=L/2)$. 
As a result, e.g., near the vortex at $(x_0=0, x_1=0)$ we expect the index to be $1$. Our goal is to show that the integral of the divergence of the lattice GHC in a region around the vortex equals the index.

\begin{table}[h!]
\centering
\begin{tabular}{ |c|c|c|c| } 
 \hline
  & \multicolumn{2}{c|}{ Wilson-like operator $\mathcal{D}_1$} & Wilson operator $\mathcal{D}_2$ \\
 \hline
 Tuned domain walls? & \multirow{2}{*}{Yes} & \multirow{2}{*}{No} & \multirow{2}{*}{No} \\ 
\cline{1-1}
 Exact zeromode? &  & & \\ 
 \hline
 Singular values & \multirow{2}{*}{---} & \multirow{4}{*}{Similar to $\mathcal{D}_2$} & Fig.~\ref{fig:singular_values_wilson_dirac} and Fig.~\ref{fig:plane_wave_wilson_dirac} for $\phi_2(x_0) \to 0$ \\
 \cline{1-1} \cline{4-4}
 Singular vectors &  & & Fig.~\ref{fig:zero_modes_wilson_dirac} and Fig.~\ref{fig:zero_modes_slices_wilson_dirac} \\
 \cline{1-2} \cline{4-4}
 Generalized Hall Current & Fig.~\ref{fig:div_exact} and Fig.~\ref{fig:div_exact_slices} &   & Fig.~\ref{fig:div_wd} and Fig.~\ref{fig:wd_div_exact_slices} \\
 \cline{1-2} \cline{4-4}
 ``Index'' & Fig.~\ref{fig:index_wilson_like} & & Fig.~\ref{fig:index_wd_vs_M} and Fig.~\ref{fig:index_wd} \\ 
 \hline
\end{tabular}
\caption{List of figures for different numerical setups presented in the paper.}
\label{table:1}
\end{table}

\subsubsection{Zeromode numerics and singular value decomposition (SVD)}
\label{subnum}

In this subsection we study the eigenvalues of the finite volume lattice operators numerically. Our goal is to map the lowest eigenstate of the suitable finite volume lattice operator to the zeromode of the infinite volume continuum fermion operator. As stated earlier, this mapping cannot be performed smoothly in the infinite volume limit by directly considering the eigenvalues of $\mathcal{D}_{1/2}$ and  $\mathcal{D}_{1/2}^{\dagger}$. Instead, we need to consider the eigenvalues of $\mathcal{D}_{1/2}^{\dagger}\mathcal{D}_{1/2}$
and $\mathcal{D}_{1/2}\mathcal{D}_{1/2}^{\dagger}$. Our goal therefore, is to find the lowest eigenvalues of $\mathcal{D}_{1/2}^{\dagger}\mathcal{D}_{1/2}$
and $\mathcal{D}_{1/2}\mathcal{D}_{1/2}^{\dagger}$ and confirm that they go to zero in the infinite volume limit. 
This discussion is organized as follows: first, we present numerical methods for finding the zeromodes of $\mathcal{D}_{1/2}\mathcal{D}_{1/2}^{\dagger}$ and $\mathcal{D}_{1/2}^{\dagger}\mathcal{D}_{1/2}$. We first apply this method to study a $0+1$ dimensional Wilson fermion operator with domain wall. We then apply it to the lattice fermion operators we wish to study in $1+1$ dimensions, i.e. $\mathcal{D}_1$ and $\mathcal{D}_2$. 

To describe the numerical technique, we use a fermion operator $\mathcal{D}$ which would serve as a proxy for both $\mathcal{D}_{1/2}$. We can now consider the spectrum of the operators $\mathcal{D}\mathcal{D}^\dagger$ and $\mathcal{D}^\dagger\mathcal{D}$ using the eigenvalue equation
\begin{eqnarray}
\mathcal{D}\mathcal{D}^\dagger u_i = \sigma_i^2 u_i,\\
\mathcal{D}^\dagger\mathcal{D} v_i = \sigma_i^2 v_i,
\end{eqnarray}
where $\sigma_i^2$ is an eigenvalue. The eigenvectors $u_i$ and $v_i$ are called left and right \textit{singular vectors} and corresponding $\sigma_i \geq 0$ is called a \textit{singular value} of $\mathcal{D}$. Note that the vectors $u_i$ and $v_i$ are linearly independent since the fermion operator is not normal, i.e. $[\mathcal{D}^\dagger,\mathcal{D}] \neq 0$.

Another possible way to arrive at the same is to look for a vector $v^\prime$ which will minimize the norm $\vert \mathcal{D} v^\prime \vert$. The square of this norm is positive-definite quadratic form given by $\mathcal{D}^\dagger\mathcal{D}$, therefore the minimum is delivered by eigenvector $v_{min}$ corresponding to the smallest eigenvalue $\sigma^2_{min}$. Analogously, $u_{min}$ will deliver the minimum of $\vert \mathcal{D}^\dagger u^\prime \vert$.

Interestingly, there is a simple relationship between $u_i$ and $v_i$, since they together with $\sigma_i \geq 0$ define a singular values decomposition (SVD) of the operator $\mathcal{D}$:
\begin{eqnarray}
\label{eq:svd}
\mathcal{D}^\dagger u_i = \sigma_i v_i,\\ \nonumber
\mathcal{D} v_i = \sigma_i u_i.
\end{eqnarray}
The SVD can be written in a compact matrix form as follows:
\begin{eqnarray}
\label{eq:svd_matrix_form}
\mathcal{D}  = U \Sigma V^\dagger,
\end{eqnarray}
where unitary matrix $U$ is composed of (column) singular vectors $u_i$, unitary matrix $V$ -- of singular vectors $v_i$ and $\Sigma$ is a diagonal matrix of corresponding singular values $\sigma_i$. 
It is clear from SVD that neither $u_i$ nor $v_i$ are straightforwardly related to eigenvectors of $\mathcal{D}$ if the operator is not normal.  However, the singular values of the operator $\mathcal{D}$ and $\mathcal{D}^{\dagger}$ map one to one to the eigenvalues of the operators $\mathcal{D}^{\dagger}\mathcal{D}$ and $\mathcal{D}\mathcal{D}^{\dagger}$. Therefore, the SVD of $\mathcal{D}/\mathcal{D}^{\dagger}$ is equivalent to eigen-decomposition of $\mathcal{D}^{\dagger}\mathcal{D}/\mathcal{D}\mathcal{D}^{\dagger}$ etc. In the rest of the paper we will refer to the lowest eigenmode of $\mathcal{D}^{\dagger}\mathcal{D}/\mathcal{D}\mathcal{D}^{\dagger}$ as near-zeromode of the operator $\mathcal{D}/\mathcal{D}^{\dagger}$ and the vectors $u_i, v_i$
as singular vectors. 

\paragraph{Wilson fermion operator in $0+1$ dimension:}
\hfill\break

We first demonstrate the utility of our approach in the simple case of $0+1$ dimensional Wilson fermion operator $\mathcal{D}_{1d}$ in the presence of a domain wall. We use periodic boundary conditions on $S^1$ and a domain wall in the fermion mass:
\begin{eqnarray}
\label{eq:1d_domain_wall}
m(x)&=&\begin{cases}
      m_+ & L/2>x\geq 0\\
      m_- & -L/2\leq x<0\\
    \end{cases}.
\end{eqnarray}

The equation of motion is given by:
\begin{eqnarray}
\mathcal{D}_{1d} \psi(x) = \frac{1}{2}\left(\psi(x+1) - \psi(x-1)\right) + m(x) \psi(x) +\frac{R}{2}\left(\psi(x+1) + \psi(x-1)-2 \psi(x)\right) = 0,
\end{eqnarray}
which in the case of $R=1$ can be simplified to:
\begin{eqnarray}
\label{eq:1d_wd_op_r_1}
\mathcal{D}_{1d} \psi(x) = (m(x) - 1) \psi(x) + \psi(x+1) = 0.
\end{eqnarray}

We numerically find singular vectors $u_i$ and $v_i$ together with singular values $\sigma_i$ of this operator and study their dependence on the lattice size $L$.

\begin{figure}[h]
    \centering
    \includegraphics[width=0.5\textwidth]{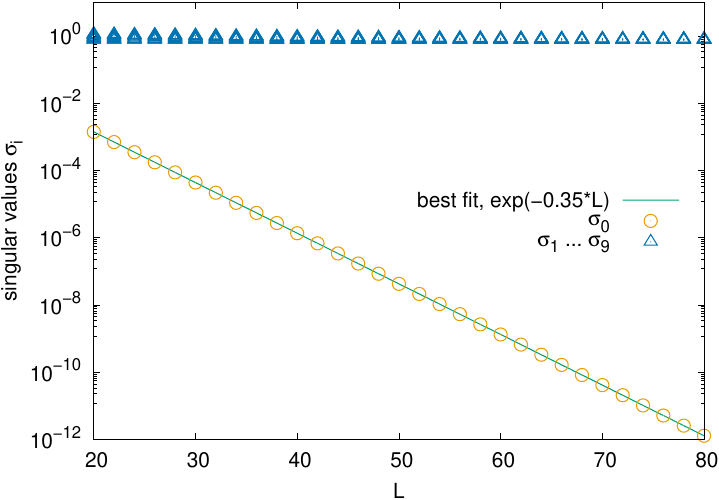}
    \caption{The flow of $10$ smallest singular values $\sigma_i$ of the $0+1$-dimensional Wilson fermion operator $\mathcal{D}_{1d}$ (Eq.~\ref{eq:1d_wd_op_r_1}) as a function of the lattice size $L$. One can clearly see that the smallest singular value follows exponential law $\sigma_{0} \sim e^{-\alpha L}$. Corresponding singular vector is localized on one of the the domain walls. The domain wall profile is given by Eq.~\ref{eq:1d_domain_wall} with $m_+ = -1$ and $m_- = 3/4$. Note logarithmic scale.}
    \label{fig:svd_1d}
\end{figure}

Let us first consider singular values $\sigma(L)$ which we depicted on the Fig.~\ref{fig:svd_1d}. We observe that the smallest singular value $\sigma_{0}$ approaches zero exponentially fast: $\sigma_{0} \sim O(e^{-L})$, whereas other singular values remain finite. This indicates that in the infinite volume there exists a zero mode of $\mathcal{D}_{1d}$ given by infinte volume limit of corresponding singular vector $v_{min}$.

\begin{figure}[h]
\begin{subfigure}{0.49\textwidth}
    \includegraphics[width=1\textwidth]{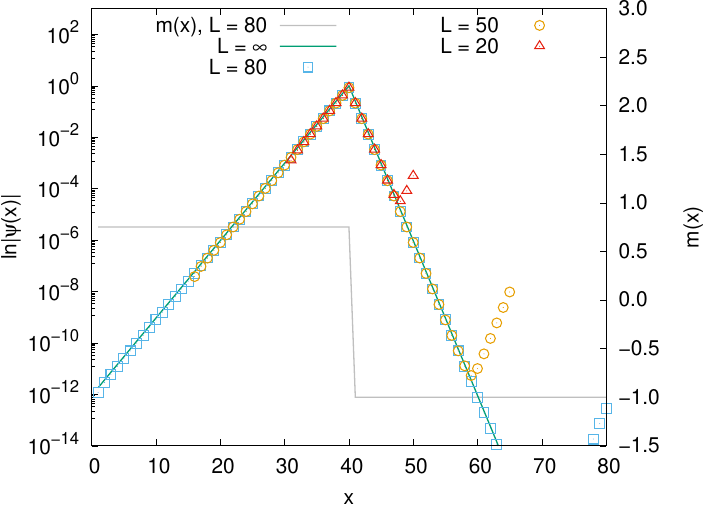} 
    \caption{Volume dependence}
    \label{fig:svd_1d_near_zero}
\end{subfigure}
\begin{subfigure}{0.49\textwidth}
    \includegraphics[width=1\textwidth]{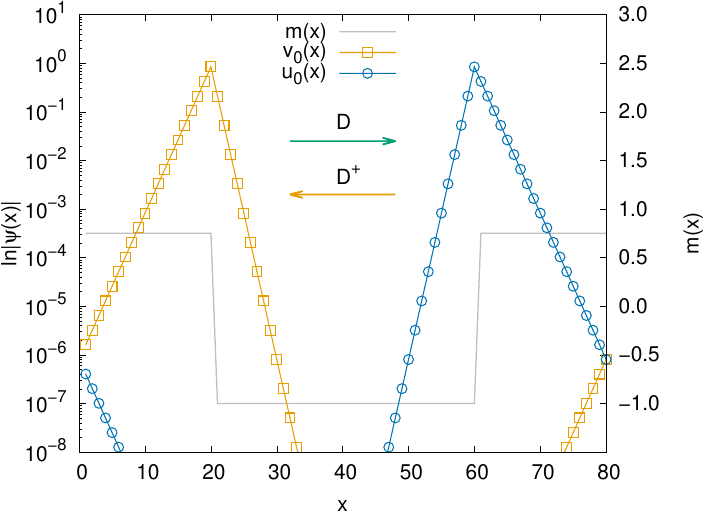} 
    \caption{Near-zeromodes of $\mathcal{D}_{1d}$ and $\mathcal{D}^\dagger_{1d}$. We show how acting on $v_0$ with $\mathcal{D}=\mathcal{D}_{1d}$ takes $v_0$ to $u_0$ and acting on $u_0$ with $\mathcal{D}_{1d}^{\dagger}$ takes $u_0$ to $v_0$.}
    \label{fig:svd_1d_near_zero_L80}
\end{subfigure}
\caption{Near-zero modes of $\mathcal{D}_{1d}$ (Eq.~\ref{eq:1d_wd_op_r_1}) in $0+1$ dimensions with $m_+=-1, m_-=3/4$.}
\end{figure}

We show the near-zero mode $v_{0}$ on the Fig.~\ref{fig:svd_1d_near_zero}. We compare it to exact solution of equation $\mathcal{D}_{1d}\psi_{inf}(x) = 0$ in the infinite volume which is given by:
\begin{eqnarray}
\label{eq:01_case_zm}
\psi_0^{inf}(x) =
\begin{cases}
\left( {1-m_{+}} \right)^{x}, &\qquad x \geq 0,\\
\left( {1-m_{-}} \right)^{x}, &\qquad x < 0,
\end{cases}
\end{eqnarray}
where $m_{\pm}$ are bulk fermion masses on either sides of the domain wall. Here we work with $m_-=3/4, m_+=-1$.
We find an excellent agreement between $v_{0}$ and $\psi_{inf}$ already for lattice sizes $L > 20$. We also show on the Fig.~\ref{fig:svd_1d_near_zero_L80} how near-zeromodes of $\mathcal{D}_{1d}$ and $\mathcal{D}_{1d}^\dagger$ are related by SVD in Eq.~\ref{eq:svd}.\\ 

\begin{figure}[h]
    \centering
    \includegraphics[width=0.8\textwidth]{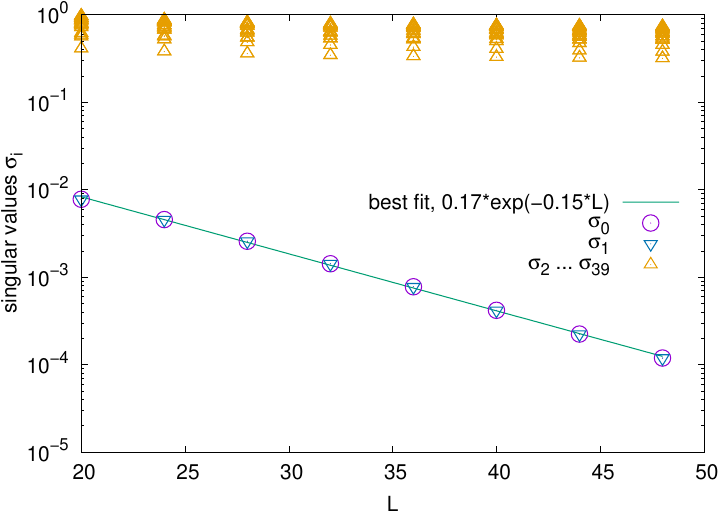} 
    \caption{The flow of $40$ smallest singular values $\sigma_i$ of the Wilson fermion operator $\mathcal{D}_2$ as a function of the lattice size $L \times L$. One can clearly see two degenerate singular values which follow exponential law $\sigma_{i=0,1} \sim e^{-\alpha L}$. Corresponding singular vectors are localized on two different vortices with winding numbers $\nu_\theta = -1$. The domain wall profile is given by Eq.~\ref{eq:domain_wall_generic} with $m_+ = \mu_+ = 1/2$ and $m_- = \mu_- = -1/2$. Note logarithmic scale.}
    \label{fig:singular_values_wilson_dirac}
\end{figure}

\begin{figure}[h]
    \centering
    \includegraphics[width=0.8\textwidth]{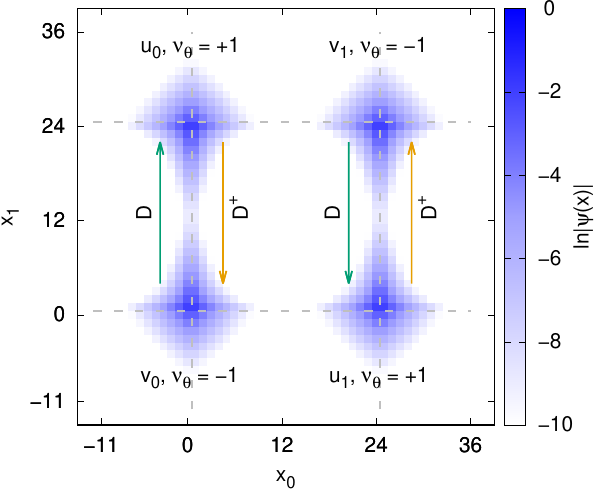} 
    \caption{Density plot of absolute values of near-zeromodes $v_0$ and $v_1$ of the Wilson fermion operator $\mathcal{D}_2$ and $u_0$ and $u_1$ of $\mathcal{D}_2^\dagger$ corresponding to two smallest singular values $\sigma_{i=0,1}$ (which are degenerate) for the lattice size $48 \times 48$. Arrows illustrate how $u_{i=0,1}$ and $v_{i=0,1}$ are related by the SVD Eq.~\ref{eq:svd}. The domain wall profile is given by Eq.~\ref{eq:domain_wall_generic} with $m_+ = \mu_+ = 1/2$ and $m_- = \mu_- = -1/2$. Gray dashed line marks the position of domain walls. Note logarithmic scale.}
    \label{fig:zero_modes_wilson_dirac}
\end{figure}

\begin{figure}[h]
    \centering
    \includegraphics[width=0.8\textwidth]{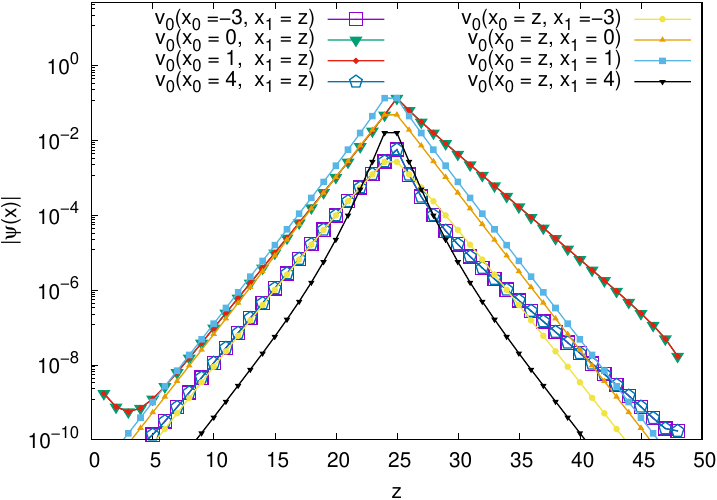} 
    \caption{Absolute value of the near-zeromode $v_0$ of the Wilson fermion operator $\mathcal{D}_2$ along several $x_0 = \mathrm{const}$ and $x_1 = \mathrm{const}$ slices for the lattice size $48 \times 48$. The domain wall profile is given by Eq.~\ref{eq:domain_wall_generic} with $m_+ = \mu_+ = 1/2$ and $m_- = \mu_- = -1/2$. Note logarithmic scale.}
    \label{fig:zero_modes_slices_wilson_dirac}
\end{figure}

\begin{figure}[h]
    \centering
    \includegraphics[width=0.8\textwidth]{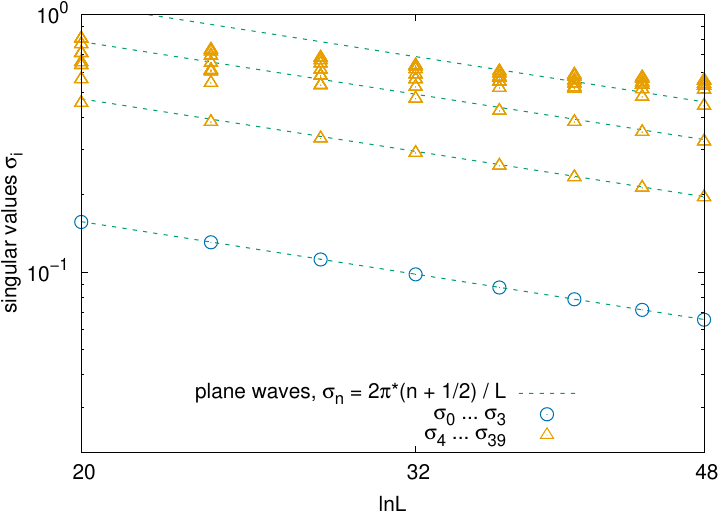} 
    \caption{The flow of $40$ smallest singular values $\sigma_i$ of the Wilson fermion operator $\mathcal{D}_2$ as a function of the lattice size $L \times L$ for very small diagnostic field $\phi_2(x_0)$ and anti-periodic boundary conditions in the direction $\mu=0$. One can clearly see that lowest singular values follow the law $\sigma_{i} \sim 2 \pi \left(i + 1/2\right) / L$ characteristic to plane waves in anti-periodic box. The domain wall profile is given by Eq.~\ref{eq:domain_wall_generic} with $-m_- = m_+ = 1/2$ and $-\mu_- = \mu_+ = 5*10^{-5}$. Note log-log scale.}
    \label{fig:plane_wave_wilson_dirac}
\end{figure}
\paragraph{Fermion operators in $1+1$ dimension:}\hfill\break

Let us now consider $1+1$ dimensional fermion operators we proposed in section \ref{lat} and analyze the corresponding zeromodes and near-zeromodes. 
As mentioned before, in $1+1$ D, it is possible to obtain an exact zeromode for the Wilson-like operator in finite volume by tuning the domain wall heights. However, we didn't find such a solution for the Wilson fermion operator. Here we will use SVD to instead find near-zeromodes for the Wilson fermion $\mathcal{D}_2$ and Wilson-like fermion operators $\mathcal{D}_1$.  The results for the Wilson-like case are very similar to the Wilson fermion case. Therefore, we only present results for the Wilson fermion case here.

In order to study the singular values of the Wilson fermion operator we use two-dimensional lattice of the size $L \times L$ and impose periodic boundary conditions. We also use the  domain wall configuration Eq.~\ref{eq:domain_wall_generic} with $0>-m_-=m_+>0$ and $0>-\mu_-=\mu_+>0$.

By performing SVD numerically for different lattice sizes $L$ we find a complete set of singular values $\sigma_i(L)$ and corresponding singular vectors $v_i(L)$ and $u_i(L)$.
Let us first consider few lowest singular values $\sigma_i(L)$ which are presented on the Fig.~\ref{fig:singular_values_wilson_dirac}. We observe that the smallest two of them (take them to be $i=0, 1$) are degenerate and exhibit clear exponential decay as $L \rightarrow \infty$. Thus, we find the first evidence for the emergence of two degenerate zero modes of the Wilson fermion operator in the infinite volume.

Let us now study corresponding singular vectors $v_i(L)$ and $u_i(L)$. Note that there are two degenerate singular vectors $v_{i=0,1}(L)$ corresponding to the lowest $\sigma_{0}=\sigma_1$. The same is true for $u_i$.  These degenerate vectors are some superposition of two near-zero modes localized on appropriate vortex defects, i.e. $v_{i=0,1}$ are superpositions of near-zeromodes on defects with winding $-1$. These two defects are localized at $(x_0=0, x_1=0)$ and $(x_0=L/2, x_1=L/2)$.
Similarly, $u_{i=0,1}$ are superpositions of near-zeromodes located on defects with winding $1$, $(x_0=0, x_1=L/2)$ and $(x_0=L/2, x_1=0)$.

At this point we can change basis by writing $v^\prime_i = \alpha_i v_0 + \beta_i v_1$ with $|\alpha_i|^2 + |\beta_i|^2 = 1$ with $i=0, 1$, in order to find near-zeromodes which are completely localized on the vortices. One can achieve this by minimizing Inverse Participation Ratio (IPR) which can serve as a measure of the localization:
\begin{eqnarray}
\mathrm{IPR} = \frac{1}{\sum_{x_0,x_1} |v^\prime(x_0,x_1)|^2}.
\end{eqnarray}
Intuitively, if a mode is uniformly distributed over entire lattice of volume $V$ then one would find that $\mathrm{IPR} = V$. On the other hand, if the mode is localized at a single point then $\mathrm{IPR} = 1$. 

Using this method we find two vectors $v'_{i=0,1}(L)$ which are exponentially localized on two vortices of the same winding number $\nu_\theta = -1$, as shown on the Fig.~\ref{fig:zero_modes_wilson_dirac} and Fig.~\ref{fig:zero_modes_slices_wilson_dirac}. Thus, we have identified two near-zermodes of the Wilson fermion operator $\mathcal{D}_2$. For convenience, we will refer to these vectors $v_{i=0,1}$ and forego the superscript prime, as in $v'\rightarrow v$. We do the same for the vectors $u_{0/1}$.
The same procedure yields two vectors $u_i(L)$ corresponding to the same two singular values localized on the other two vortices of winding number $\nu_\theta = +1$ (at $x_0=0, x_1=L/2$ and $x_0=L/2, x_1=0$).

Finally, let us describe how near-zeromodes behave if one switches the diagnostic field off, i.e. $\phi_2 \rightarrow 0$. If the lattice volume is kept fixed, then at sufficiently small $\phi_2$ the near-zeromodes completely delocalize in the direction $\mu=0$, and the SVD spectrum become consistent with that of $\phi_2 = 0$ case. Namely, we find that near-zeromodes transform into plane wave excitations living on the two remaining domain walls. This can be seen by direct inspection of $\vert v_i(x_0,x_1) \vert$ and from the behavior of singular values $\sigma_i(L) \sim 2\pi n / L$ characteristic to the spectrum of plane waves in the finite box. Furthermore, lowest singular values are $4$ times degenerate accounting for $2$ remaning domain walls and $2$ possible spinor polarizations. Additionally, by imposing anti-periodic boundary condition in the $\mu=0$ direction we again observe that the flow of singular values $\sigma_i(L) \sim 2\pi (n + 1/2) / L$ is characteristic to that of plane waves in the anti-periodic box, see Fig.~\ref{fig:plane_wave_wilson_dirac}. The true near-zeromode should not, in general, be sensitive to such change of boundary conditions.
This reorganization happens because for sufficiently small $\phi_2$ the localization width of the near-zero modes become comparable or bigger than the lattice size, thus it completely delocalizes. If $\phi_2$ is kept fixed then one should recover the near-zeromodes by increasing the volume. Therefore we find that limits $\phi_2 \rightarrow 0$ and $L \rightarrow \infty$ do not commute. In order to correctly define the ``index" from the finite volume analysis one has to take infinite volume limit first and only then switch the diagnostic field off.

\section{Generalized Hall Current in the finite volume}
In this part we will study the realization of the Generalized Hall Current (GHC) for the Wilson-like and the Wilson fermions and corresponding ``indices". Before we proceed to the computations, let us outline the plan of this section. First, we will present how we've computed the GHC on the lattice. Next, we will study GHC for the Wilson-like operator $\mathcal{D}_1$, taking the domain wall heights to satisfy the tuning condition (Eq.~\ref{eq:tuning_wilson_like_pbc}). This will illustrate how the GHC reproduces the index of the fermion operator in finite volume for the case when there is an exact zeromode.
 This will give us an opportunity to study GHC and its relation to the index without complications of the finite volume effects.
 
 Next, we will proceed to study of Wilson fermion operator and see how near-zeromodes and finite volume effects influence the realization of the GHC. Results for the Wilson-like operator in the same setup (when exact zeromodes are absent) are essentially the same, therefore we will not present them.

\subsection{Computation of the Generalized Hall Current on the lattice}

The lattice generalized Hall current $J^H_\mu(x)$ can be defined as follows:
\begin{eqnarray}
\label{eq:lat_ghc}
J^H_\mu(x) = \bar{\Psi} \tilde{\Gamma}_\mu(x) \Gamma_\chi \Psi
\end{eqnarray}
where $\tilde{\Gamma}_\mu(x)$ is given by:
\begin{eqnarray}
\label{eq:lat_gamma_mu}
\tilde{\Gamma}_\mu(x) = -i \left. \frac{\delta K(A_\mu(x))}{\delta A_\mu(x)} \right|_{A_\mu(x) = 0}.
\end{eqnarray}

Here $A_\mu(x)$ is a $U(1)$ gauge field and $K(A_\mu(x))$ is a gauged lattice Dirac operator of the double theory obtained via standard Peierls substitution $\delta_{x+a_\mu,y} \rightarrow \delta_{x+a_\mu,y} \, \exp(i A_\mu(x))$. 

The expectation value of $J^H_\mu(x)$ is evaluated numerically by straightforward computation of the matrix $(K + M)^{-1}$ and taking a trace. The divergence is computed as usual with the help of lattice backward difference $\nabla^B_\mu$:
\begin{eqnarray}
\label{eq:lat_div}
\nabla^B_\mu J^H_\mu(x) = \sum_{\mu=0,1} \left( J^H_\mu (x - a_\mu) - J^H_\mu(x) \right).
\end{eqnarray}

We posit that the space-time integral of the divergence should produce the ``index" of interest.
We compute the ``index" $I_{lat}$ according to the lattice version of the Eq.~\ref{eq:index_vs_divj}:
\begin{eqnarray}
\label{eq:lat_index}
I_{lat} = -\frac{1}{2}\sum_{x \in S} \nabla^B_\mu J^H_\mu(x)
\end{eqnarray}
where $S$ is the area over which the divergence of the lattice GHC current $J^H_\mu(x)$ is integrated. The area $S$ can be an entire lattice, however in that case the total index has to vanish. Thus we will integrate only over some portion of the lattice adjacent to the defect (vortex) of interest. To implement this, we divide the lattice into $4$ equal squares centered around each of the $4$ vortices created by the domain walls and then integrate the divergence of lattice GHC on these four squares separately to compute the corresponding index. 

\subsection{GHC for Wilson-like lattice operator and exact zeromodes}

Let us first present results for GHC for Wilson-like operator $\mathcal{D}_1$ when domain wall configuration satisfies the tuning condition Eq.~\ref{eq:tuning_wilson_like_pbc}. In this case there is an exact zeromode for the fermion operator in finite volume.

\begin{figure}[h]
\begin{subfigure}{0.49\textwidth}
\includegraphics[width=0.9\linewidth]{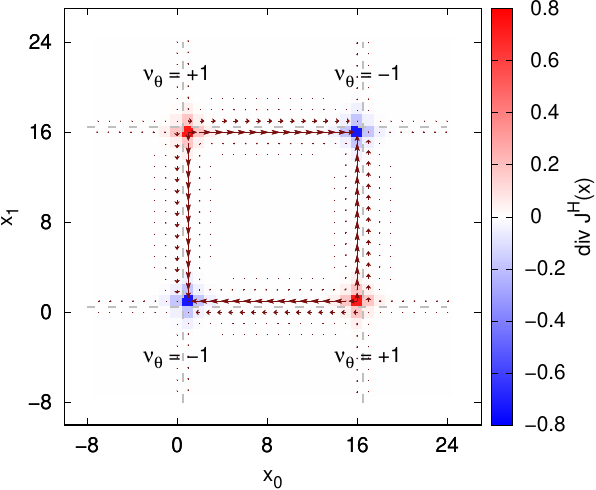}
\caption{Wilson-like operator $\mathcal{D}_1$, domain wall is $m_- = \mu_- = -1$ and $m_+ = \mu_+ = 1/2$ which satisfies Eq.~\ref{eq:tuning_wilson_like_pbc}.}
\label{fig:div_exact}
\end{subfigure}
\begin{subfigure}{0.49\textwidth}
\includegraphics[width=0.9\linewidth]{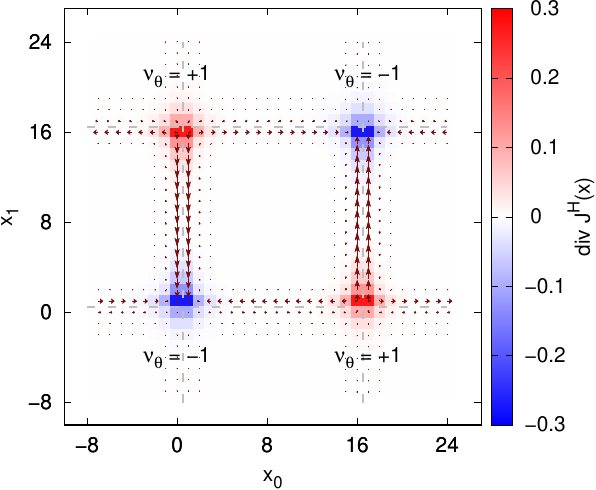}
\caption{Wilson operator $\mathcal{D}_2$, domain wall is $m_- = \mu_- = -1/2$ and $m_+ = \mu_+ = 1/2$.}
\label{fig:div_wd}
\end{subfigure}
\caption{The lattice GHC $J^H_\mu(x)$ and its divergence on the lattice $32 \times 32$ for (a) Wilson-like operator $\mathcal{D}_1$ with $M=10^{-5}$ and (b) Wilson operator $\mathcal{D}_2$ with $M=0.21$. Arrows represent the current $J^H_\mu(x)$ with their size being proportional to the current magnitude. Gray dashed lines show the location of domain walls.}
\label{fig:ghc_div}
\end{figure}

\begin{figure}[h]
\begin{subfigure}{0.48\textwidth}
\includegraphics[width=0.9\linewidth]{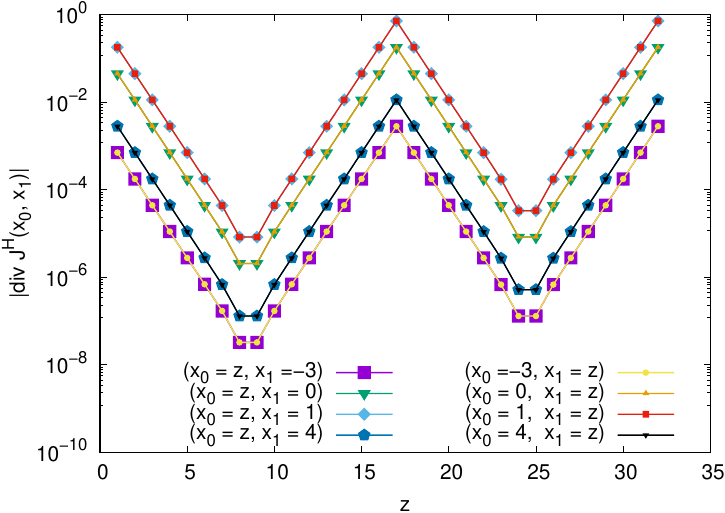}
\caption{Wilson-like operator $\mathcal{D}_1$.}
\label{fig:div_exact_slices}
\end{subfigure}
\begin{subfigure}{0.48\textwidth}
\includegraphics[width=0.9\linewidth]{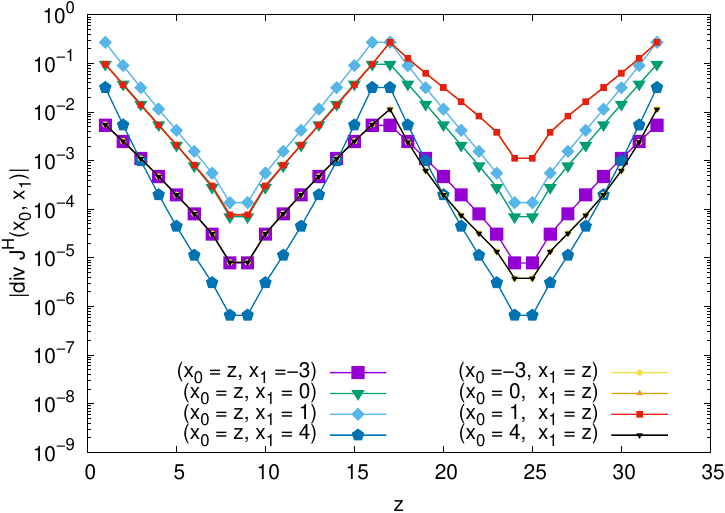}
\caption{Wilson operator $\mathcal{D}_2$.}
\label{fig:wd_div_exact_slices}
\end{subfigure}
\caption{Several slices of the (log of) divergence of $J^H_\mu(x)$ for (a) Wilson-like operator $\mathcal{D}_1$ with $M=10^{-5}$ and (b) Wilson operator $\mathcal{D}_2$ with $M=0.21$.}
\label{fig:div_slices}
\end{figure}

We've computed the GHC $J^H_\mu(M)$ and the ``index" $I_{lat}(M)$ for several values of the regulator mass from $M =10^{-5}$ to $2$ on the lattice $L \times L = 32 \time 32$. We present the current $J^H_\mu(x)$ and its divergence on the Fig.~\ref{fig:div_exact} for the smallest value of $M$, with $M = 10^{-5}$. We observe that the divergence is localized around the vortices. It has maximal value at the vortex center. The sign is consistent with the winding number of the defect. The current $J^H_\mu(M)$ flows preferably along the edges of the domains from one vortex to another.
The divergence exhibits an exponential decay around the vortex as shown on Fig.~\ref{fig:div_exact_slices}.

\begin{figure}[h]
\begin{subfigure}{0.48\textwidth}
\includegraphics[width=0.9\linewidth]{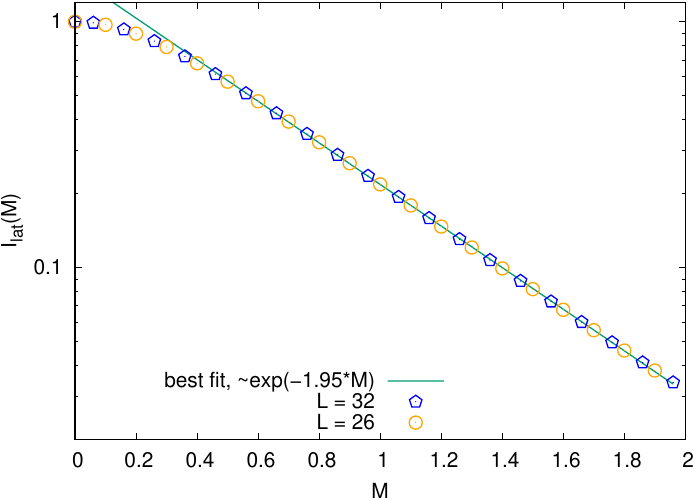}
\caption{Lattice ``index" as a function of $M$ for several values of $L$.}
\label{fig:index_wilson_like_vs_M}
\end{subfigure}
\begin{subfigure}{0.48\textwidth}
\includegraphics[width=0.9\linewidth]{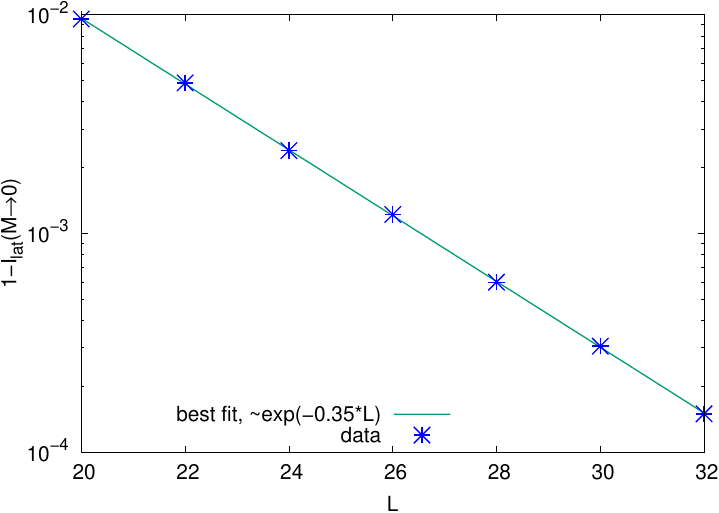}
\caption{Convergence of the lattice ``index" to the infinite volume value $\pm 1$.}
\label{fig:index_wilson_like_vs_L}
\end{subfigure}
\caption{Dependence of the lattice ``index" $I_{lat}(M)$ on $M$ and lattice size $L$ for the Wilson-like operator $\mathcal{D}_1$ in the presence of exact zeromode. The domain wall configuration is given by Eq.~\ref{eq:domain_wall_generic} with $m_- = \mu_- = -1$ and $m_+ = \mu_+ = 1/2$ which satisfies Eq.~\ref{eq:tuning_wilson_like_pbc}. Compare this figure to Fig.~\ref{fig:index_wd_vs_M}.}
\label{fig:index_wilson_like}
\end{figure}

Now we want to verify that the space-time integral of the divergence of the lattice GHC produces the correct ``index". As discussed previously, we divided the lattice into 4 equal squares centered around each vortex and performed integration of the divergence of GHC over them. Due to the exponential decay of the GHC away from the defect, we expect that that the integral would approach infinite volume value quickly. The resulting ``index" $I_{lat}(M)$ is shown in the Fig.~\ref{fig:index_wilson_like_vs_M} as function of $M$. We observe that it clearly goes towards $\pm 1$ as $M \rightarrow 0$. The sign of the index depends on the vortex defect in consideration. Also, as expected, for very large $M$ the ``index" approaches zero with increasing $M$.

In order to quantify finite volume effects we have computed the deviation:
\begin{eqnarray}
\epsilon(L) = \vert \pm 1 -  I_{lat}(M \rightarrow 0)\vert
\end{eqnarray}
where the plus or minus sign is chosen according to the winding of the vortex and $I_{lat}$ is the corresponding ``index" computed by integrating the $\nabla_{\mu}^B J_{\mu}^H$. This function is shown in the Fig.~\ref{fig:index_wilson_like_vs_L} where one can see that the error is indeed exponentially small: $\epsilon(L) \sim e^{-L}$. Therefore, after performing infinite volume extrapolation our computations show that the lattice GHC correctly reproduces the index of the Euclidean fermion operator. Finally, we find that generalized hall current and divergence vanish when $\phi_2 \rightarrow 0$ for fixed $L$ and $M$. This shows that we have to take the infinite volume limit first and then take $\phi_2$ to zero in order to retain a nonzero index in the limit of $\phi_2\rightarrow 0$.

\subsection{GHC for Wilson fermion operator and near-zero modes}
We now present results for GHC and the index for the Wilson fermion operator $\mathcal{D}_2$. The results for the untuned Wilson-like operator are very similar. 

\begin{figure}[h]
    \centering
    \includegraphics[width=0.8\textwidth]{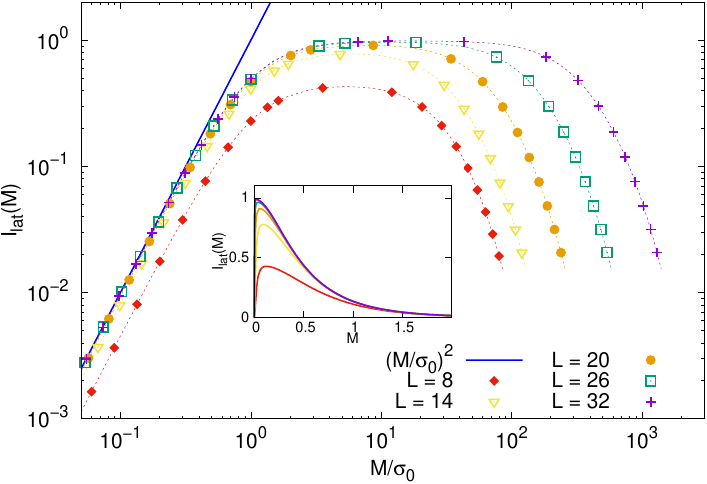} 
    \caption{The ``index" $I_{lat}(M)$ of the Wilson fermion operator $\mathcal{D}_2$ for several lattice sizes $L = 8 \dots 32$. The domain wall configuration is given by Eq.~\ref{eq:domain_wall_generic} with $m_- = \mu_- = -1/2$ and $m_+ = \mu_+ = 1/2$. Here $\sigma_0$ is the smallest singular value and the inset shows the same data but in the linear scale.}
    \label{fig:index_wd_vs_M}
\end{figure}

\begin{figure}[h]
\begin{subfigure}{0.48\textwidth}
    \includegraphics[width=0.95\textwidth]{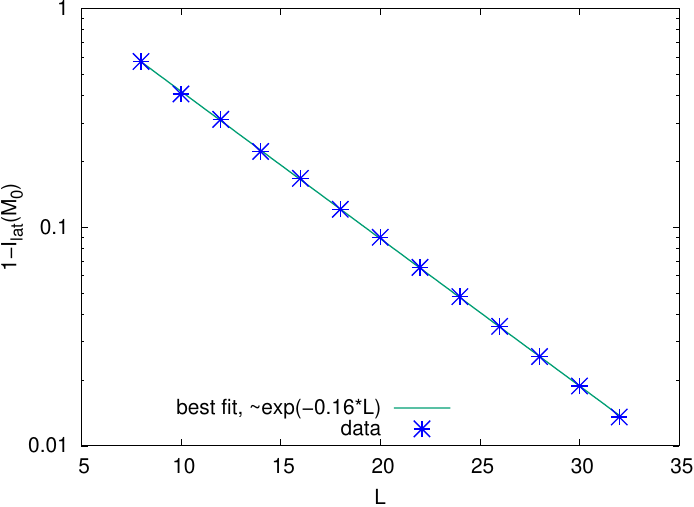} 
    \caption{Convergence of the maximal value to $\pm 1$.}
    \label{fig:index_wd_vs_L}
\end{subfigure}
\begin{subfigure}{0.48\textwidth}
    \includegraphics[width=0.95\textwidth]{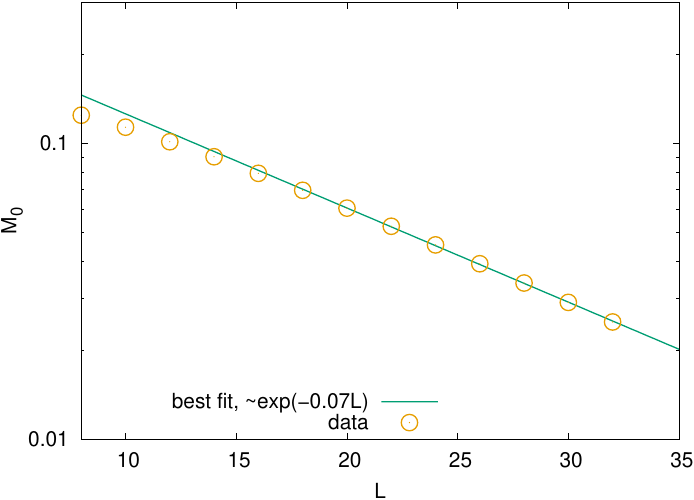} 
    \caption{The position of the maximum.}
    \label{fig:index_wd_vs_M0}
\end{subfigure}
\caption{Dependence of the value and the position of the maximum of the lattice ``index" $I_{lat}(M)$ on lattice size $L$ for the Wilson fermion operator $\mathcal{D}_2$. The domain wall configuration is given by Eq.~\ref{eq:domain_wall_generic} with $m_- = \mu_- = -1/2$ and $m_+ = \mu_+ = 1/2$.}
\label{fig:index_wd}
\end{figure}

We use the same strategy in order to compute the ``index" which is presented in the Fig.~\ref{fig:index_wd_vs_M} for several values of $M$ and lattice sizes $L = 8 \dots 32$. First of all, we observe that the ``index" vanishes when we naively take $M \rightarrow 0$. This is an expected behaviour since the spectrum of $\mathcal{D}_2$ is strictly speaking gapped: $\sigma_0 \sim \exp(-L)\neq 0$. In order to understand it better one can expand contribution of $\mathrm{Dim}(\mathrm{ker} \,\mathcal{D}_2)$ in powers of $M/\sigma_0 \ll 1$:
\begin{eqnarray}
\frac{M^2}{D_2^\dagger D_2 + M^2} = \frac{M^2}{\sigma_0^2} + O\left(\frac{M^4}{\sigma_0^4}\right).
\end{eqnarray}
We indeed find this dependence as shown on the Fig.~\ref{fig:index_wd_vs_M}. The ``index" exhibits a pronounced maximum at some $M_0 > \sigma_0$ and then
decays exponentially fast as $M \rightarrow \infty$. We find that the maximum tends to $\pm 1$ as lattice size gets bigger, also exponentially fast, as illustrated on the Fig.~\ref{fig:index_wd_vs_L}. Moreover, the position of the maximum $M_0$ tends to zero as $L \rightarrow \infty$ exponentially as well, see Fig.~\ref{fig:index_wd_vs_M0}.
Therefore we find in order to reproduce the index of the fermion operator one has to take infinite volume limit first and only then $M=M_0 \rightarrow 0$.

\section{Conclusions}
In this paper we extended the idea of generalized Hall current proposed in \cite{kaplan2021index, Kaplan:2022uoo} to discrete space-time in finite volume. Our construction is focused on one of the several examples presented in \cite{kaplan2021index, Kaplan:2022uoo}: $1+1$ dimensional Dirac fermion with a domain wall in its mass. It is well known that the domain wall hosts massless fermion in the continuum. The continuum GHC construction connects the existence of this massless fermion to a Euclidean fermion operator with an index of $1$ by turning on some diagnostic field in the theory. We extend this construction to discrete Euclidean space-time in finite volume ($S^1\times S^1$) by introducing higher dimensional operators which we call Wilson-like and Wilson terms. We tackle several nontrivial features associated with a finite volume analysis which includes the net vorticity of the defects on $S^1\times S^1$ being zero. We have four defects on the lattice, two vortices and two anti-vortices. In order to mimic the GHC construction of the continuous infinite volume space-time, we focus on the region of space-time around only one of these vortices. We were successful in engineering a nonzero index for the fermion operator on each of these vortices. We then computed the lattice GHC to show that the space-time integral of its divergence computed locally reproduced the ``index" correctly. 

Future research directions involve extending this lattice finite volume construction to higher dimensional theories. 
Ref. \cite{kaplan2021index, Kaplan:2022uoo} constructed the continuum GHC for several examples, including the $1+1$ dimensional example we focus on here. The other examples included domain wall fermions in higher dimensions. The GHC construction in these higher dimensional examples involved diagnostic background gauge fields as well as diagnostic scalar and pseudo-scalar fields. Our plan is to extend these continuum constructions to the lattice.
Also, the continuum construction of GHC in \cite{kaplan2021index, Kaplan:2022uoo} applies to free fermion theories. In particular, the GHC is computed using a one-loop Feynman diagram in perturbation theory. It is however well known that in a multiflavor theory, introducing interactions can sometimes gap out massless fermions through nonperturbative effects. This is even more interesting when the interaction in question do not break any anomalous symmetries of the non-interacting theory. E.g. see symmetric mass generation \cite{Wang:2022ucy, Wang:2013yta, You:2014vea, Wang:2018ugf, Zeng:2022grc, Tong:2021phe, Razamat:2020kyf, PhysRevX.8.011026, xu2021greens}. The non-perturbative effects of interactions on the GHC may not be captured using a one loop Feynman diagram as described in \cite{kaplan2021index, Kaplan:2022uoo}. One may need to resort to a numerical analysis to uncover these effects. 
Even though our lattice GHC construction was formulated for non-interacting $1+1$ Dimensional fermions, it can be easily modified to take into account interactions. This will enable us to compute the generalized Hall current taking into account non-perturbative effects. 
\section{Acknowledgement}
We acknowledges support from the U.S. Department of Energy,
Nuclear Physics Quantum Horizons program through the Early Career Award DE-SC0021892.
\clearpage

\printbibliography

\end{document}